\documentclass{icrc2009}

\usepackage{graphicx}   
\usepackage{caption}    
\usepackage[font=footnotesize]{subfig} 
\usepackage{fixltx2e}
\usepackage{url}

\newcommand{\shorttitle}[1]%
{\markboth{Proceedings of the 31\MakeLowercase{$^{st}$} ICRC, {\L}\'{o}d\'{z} 2009}{#1} }
\newcommand{\etal}{\MakeLowercase{\textit{et al. }}} 

\begin{document}
\title{Phase Fresnel Lens Development for X-ray and Gamma-ray Astronomy}

\author{\IEEEauthorblockN{John Krizmanic\IEEEauthorrefmark{1}\IEEEauthorrefmark{2},
			  Gerald Skinner\IEEEauthorrefmark{3}\IEEEauthorrefmark{2},                        
                           Zaven Arzoumanian\IEEEauthorrefmark{1}\IEEEauthorrefmark{2},
                           Vlad Badilita\IEEEauthorrefmark{4},\\
                           Neil Gehrels\IEEEauthorrefmark{2},
                           Keith Gendreau\IEEEauthorrefmark{2},
                           Reza Ghodssi\IEEEauthorrefmark{4},
                           Nicolas Gorius\IEEEauthorrefmark{5}\IEEEauthorrefmark{2},
                           Brian Morgan\IEEEauthorrefmark{6},\\
                           Lance Mosher\IEEEauthorrefmark{4}\IEEEauthorrefmark{8}, and
                           Robert Streitmatter\IEEEauthorrefmark{2}\IEEEauthorrefmark{9}}
                            \\
\IEEEauthorblockA{\IEEEauthorrefmark{1}CRESST/Universities Space Research Association}
\IEEEauthorblockA{\IEEEauthorrefmark{2}NASA Goddard Space Flight Center, Greenbelt, Maryland 20771 USA}
\IEEEauthorblockA{\IEEEauthorrefmark{3}CRESST/University of Maryland, College Park, Maryland 20742 USA}
\IEEEauthorblockA{\IEEEauthorrefmark{4}MEMS Sensors and Actuators Laboratory, University of Maryland, College Park, Maryland 20742 USA}
\IEEEauthorblockA{\IEEEauthorrefmark{5}Catholic University of America, Washington, DC 20064}
\IEEEauthorblockA{\IEEEauthorrefmark{6}US Army Research Laboratory, Adelphi, Maryland 20783 USA}
\IEEEauthorblockA{\IEEEauthorrefmark{7}present address University of Freiburg, Freiburg, Germany}
\IEEEauthorblockA{\IEEEauthorrefmark{8}present address Lockheed Martin Space Systems Company, Newtown, Pennsylvania 18940}
\IEEEauthorblockA{\IEEEauthorrefmark{9}retired}
\vspace{-0.5cm}
}

\shorttitle{John Krizmanic \etal PFL Development for X-ray and Gamma-ray Astronomy}
\maketitle

\begin{abstract}
In principle, diffractive optics, particularly Phase Fresnel Lenses (PFLs), offer the ability to construct large, diffraction-limited, and highly efficient X-ray/$\gamma$-ray telescopes, leading to dramatic improvement in angular resolution and photon flux sensitivity. As the diffraction limit improves with increasing photon energy, gamma-ray astronomy would offer the best angular resolution over the entire electromagnetic spectrum. A major improvement in source sensitivity would be achieved if meter-size PFLs can be constructed, as the entire area of these optics focuses photons. We have fabricated small, prototype PFLs using Micro-Electro-Mechanical Systems (MEMS) fabrication techniques at the University of Maryland and measured near diffraction-limited performance with high efficiency using 8 keV and higher energy X-rays at the GSFC 600-meter Interferometry Testbed.  A first generation, 8 keV PFL has demonstrated imaging corresponding to an angular resolution of approximately 20 milli-arcseconds with an efficiency $\sim$70$\%$ of the theoretical expectation. The results demonstrate the superior imaging potential in the X-ray/$\gamma$-ray energy band for PFL-based optics in a format that is scalable for astronomical instrumentation.  Based upon this PFL development, we have also fabricated a `proof-of-principle' refractive-diffractive achromat and initial measurements have demonstrated nearly uniform imaging performance over a large energy range.  These results indicate that the chromaticity inherent in diffractive optics can be alleviated.
 \end{abstract}

\begin{IEEEkeywords}
 Diffractive Optics, X-ray Optics, Gamma-ray Optics
 \end{IEEEkeywords}
 
 \section{Introduction}
Phase Fresnel Lenses (PFLs) have the capability for highly efficient, diffraction-limited imaging at X-ray and $\gamma$-ray energies. In principle, PFLs can be fabricated in meter or larger diameters, and a telescope employing such a PFL would have unprecedented angular resolution \cite{Skinner2001,Skinner2002}: the diffraction limit is $<$ 50 micro-arcsecond ($\mu^{\prime\prime}$) at 6 keV for a meter diameter lens. This corresponds to more than a factor of 10,000 improvement over current capabilities, such as that demonstrated by the Chandra telescope. The large effective area and small focal spot size lead to extremely good flux sensitivity, and at high energies, where no other technique offers precise focusing, the sensitivity would be unprecedented. PFLs are inherently light; a silicon PFL imaging at 100 keV has an areal density of $\sim$1 kg/m$^2$. An advanced telescope using PFLs with $<$ $1~\mu^{\prime\prime}$ angular resolution would enable imaging of regions surrounding the event horizon of the massive black holes in the nuclei of nearby galaxies. A milli-arcsecond (m$^{\prime\prime}$) X-ray/$\gamma$-ray telescope would open a new window on a wealth of astrophysical phenomena, such as imaging of the core regions of AGNs or discerning the anatomy of `hidden' AGNs \cite{Rita}.

We have fabricated PFLs at the MEMS Sensors and Actuators Laboratory (MSAL) at the University of Maryland using gray-scale technology \cite{Morgan1} to scale down the spatial size,f in order to enable characterization measurements, from that needed for astronomical instrumentation. The inherent imaging performance of this technology was characterized in a NASA/GSFC 600-meter X-ray beam line, where issues such as PFL-areal uniformity were also investigated. Two generations of PFLs have been designed and characterized. A first generation of lenses that were designed to image at 8 keV and whose characterization \cite{Krizmanic2007} led to fabrication improvements \cite{Morgan2}, leading to the fabrication of a second generation of PFLs designed to operate at 17.4 keV.  The design of the 17.4 keV PFLs formed the basis of the fabrication of the diffractive component of a contact pair achromat whose imaging performance has been measured from 8 to 11.3 keV.
 
 \section{Phase Fresnel Lenses: Design and Fabrication}
A Phase Fresnel Lens (PFL) is a circular diffraction grating with the pitch of the N concentric annuli becoming smaller in a prescribed manner with increasing radius.  The radial profile of each annulus, or full Fresnel zone, in an ideal PFL is exactly matched to the optical path needed to coherently concentrate incident radiation into the primary focus \cite{Miyamoto1961}, with the performance improving as N increases. The thickness of material is varied in each full Fresnel zone from zero to a maximum thickness of $t_{2\pi}$, the length required to obtain a $2\pi$ phase shift for the material at a specific photon energy. Thus PFLs offer performance advantages over other diffractive optics such as Zone Plates, that approximate the required lens profile within each Fresnel zone.  In an ideal PFL, all power is concentrated into the first order focus and a maximum theoretical efficiency approaching 100$\%$ is obtained. In practice, the exact Fresnel profile of a PFL is approximated by a number of steps (P) with improved performance for larger P: \cite{Dammann1970}: P=8 has an efficiency of 95$\%$, ignoring absorption.  The effects of absorption on the efficiency can be calculated to account for the material in the PFL, even for profiles that are away from the ideal \cite{Kirz1974, Krizmanic2006a}.  PFLs can be constructed with Fresnel ridge heights ($t_F$) corresponding to multiples of $t_{2\pi}$. For a 4$\pi$ phase shift, $t_F=2 t_{2\pi}$  with the Fresnel profile continuing over 2 full Fresnel zones. This effectively doubles the minimum Fresnel ridge spacing, at the outermost portion of the PFL, for a fixed diameter, at the expense of accepting a modest reduction in efficiency.

The focal length ($f$) of a PFL is related to the smallest pitch of the full Fresnel zones ($p_{min}$, located at the outermost lens radii), the diameter of the lens ($d$), and the incident photon energy ($E_{\gamma}$):
  \begin{equation}
   f = \frac{p_{min} d}{2 \lambda} = 4 \times 10^{2} \Bigl[\frac{p_{min}}{1~mm}\Bigr] \Bigl[\frac{d}{1~m}\Bigr]\Bigl[\frac{E_{\gamma}}{1~keV}\Bigr]~~km
    \label{focal_eqn}
   \end{equation}
This implies that long focal lengths are required for PFLs designed for X-ray and higher energies: the parameters appropriate for a ground-testable PFL, such as $p_{min} =10~\mu$m and $d =3$ mm, lead to a focal length of  $\approx 100$ m for a PFL designed to image at 8 keV. The focal length of a large diameter optic for an astronomical instrument necessitates placement of the lens and the photon detector on separate spacecraft in a formation-flying configuration for an eventual mission \cite{Krizmanic2006b}.
 \begin{figure}[!t]
  \centering
  \includegraphics[width=2.5in]{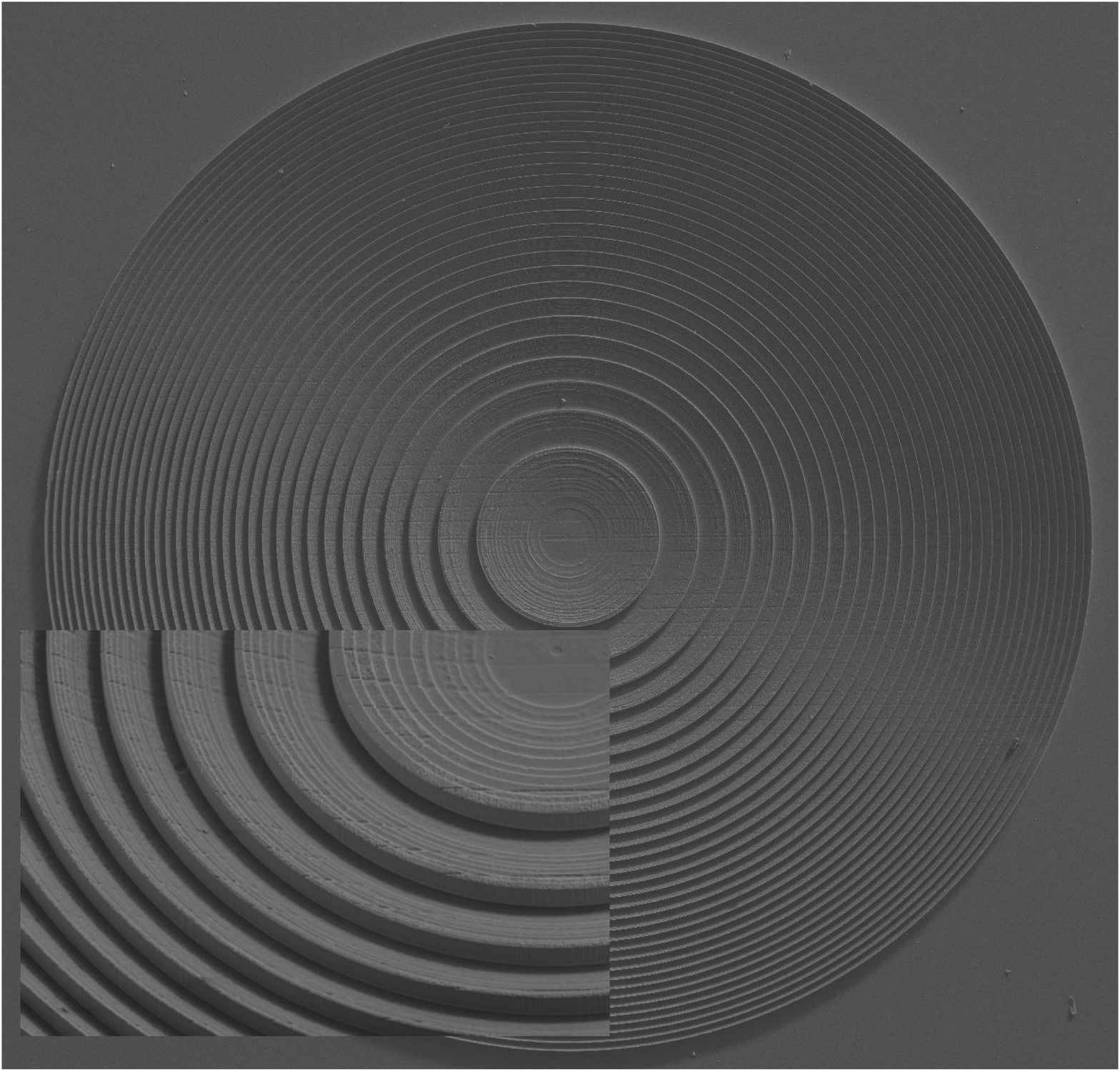}
  \caption{A SEM of a 3 mm PFL fabricated in the MSAL at the University of Maryland and characterized at 8 keV in the GSFC 600 m test beam. The inset shows a zoom of the interior Fresnel profile.}
  \label{pfl_fig}
 \end{figure}
Equation \ref{focal_eqn} details the inherent chromaticity of simple PFLs, and chromatic aberration is a dominant component to the angular resolution of a PFL \cite{Skinner2001}.
A combination refractive-diffractive lens can significantly increase the wavelength band within which near diffraction-limited angular resolution is obtained \cite{Skinner2002, Skinner2004}: the inclusion of a refractive lens with a PFL forms a contact pair achromat. If the ratio of the refractive to diffractive focal lengths is $2$ (refractive: $f_R \propto \lambda^{-2}$; PFL: $f_Z \propto \lambda^{-1}$), the first order dispersive effects cancel and the compound lens has a focal length twice that of the PFL component.  The magnitude of the attenuation due to the inclusion of the refractive component is wavelength and material dependent, with low $Z$ materials being desirable at lower X-ray energies \cite{Chantler}.

\subsection{PFL Fabrication}
\begin{figure}[!t]
  \centering
  \includegraphics[width=2.5in]{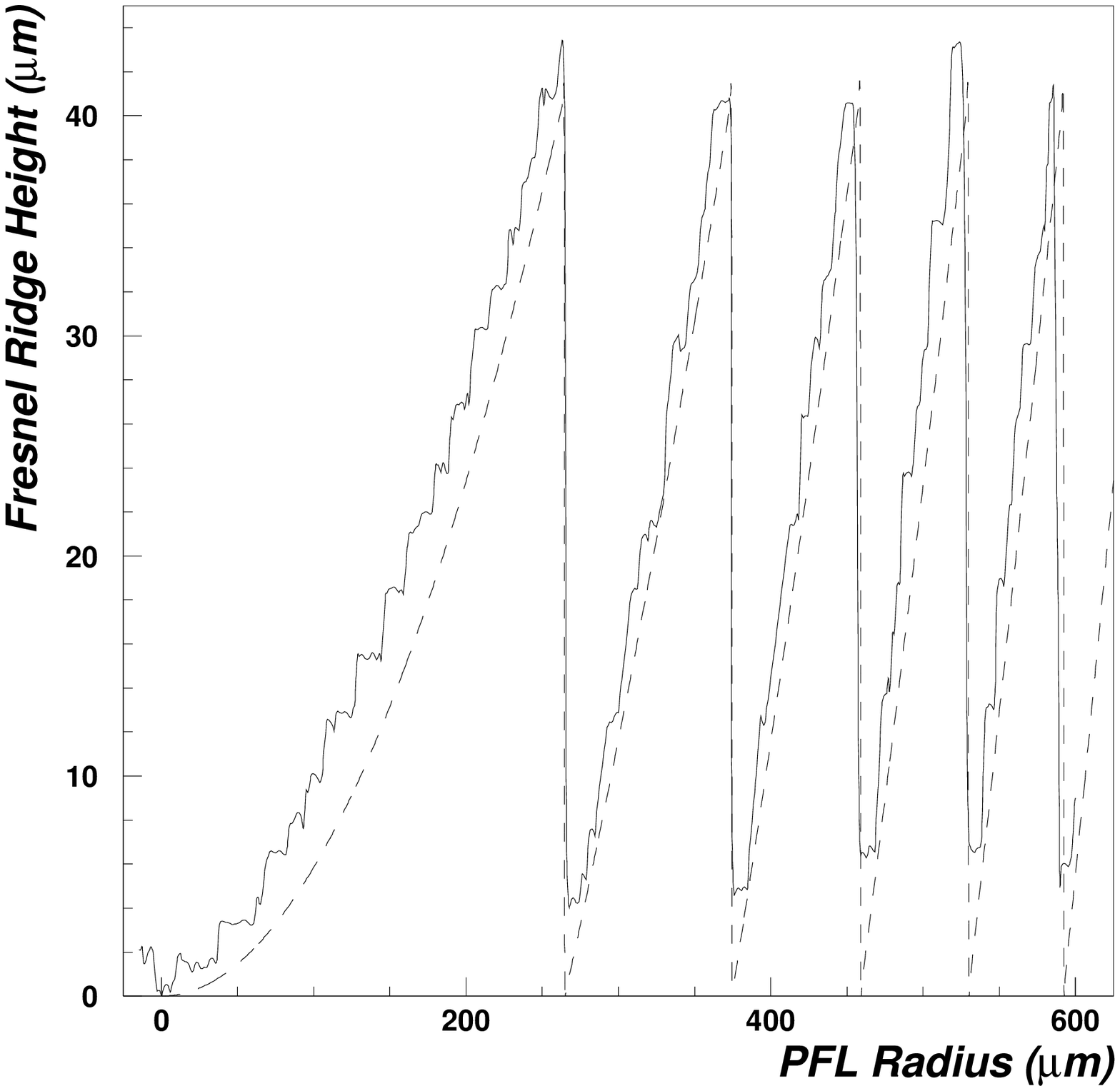}
  \includegraphics[width=2.5in]{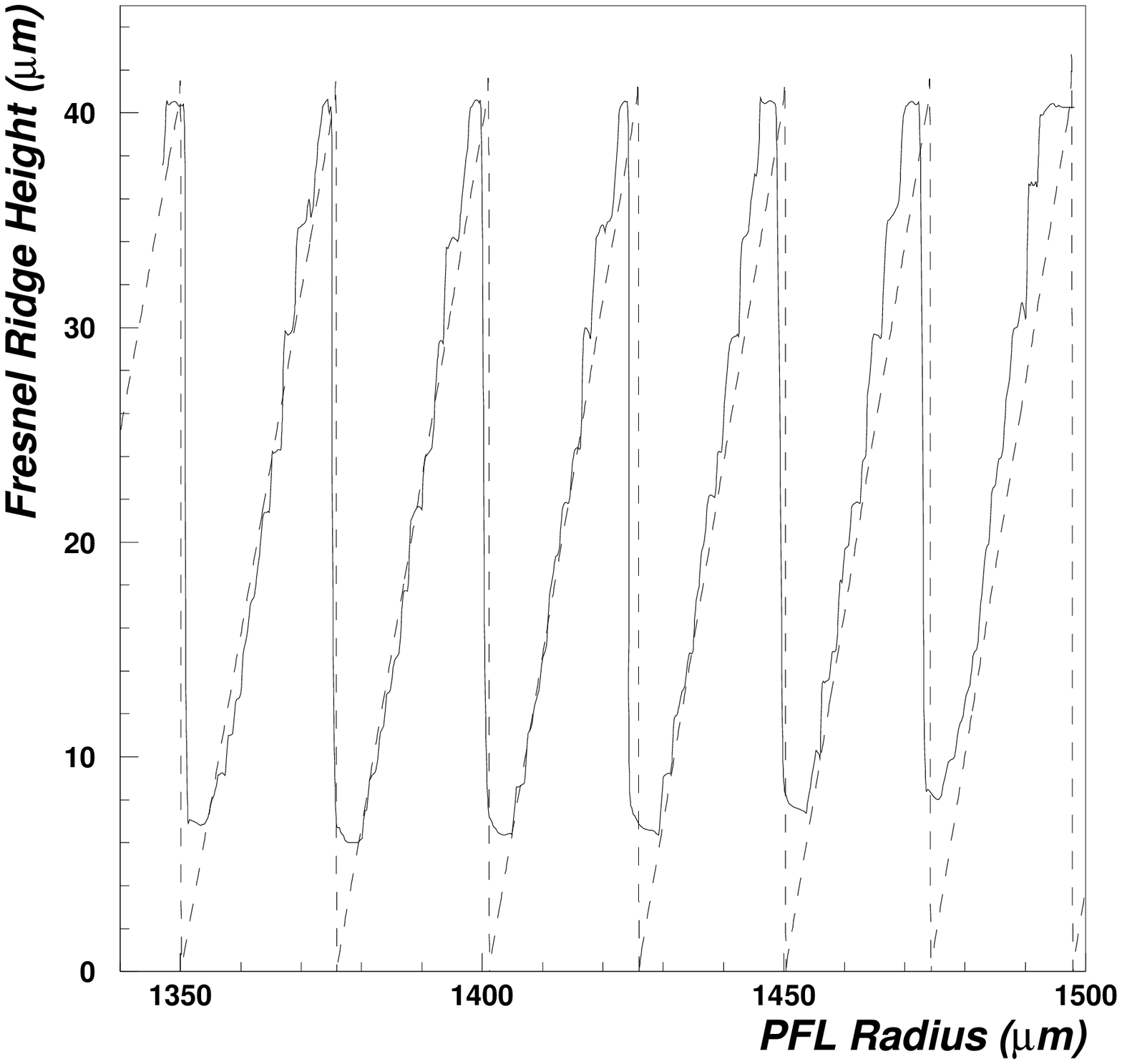}
  \caption{The measured profile (solid) of a fabricated, 3 mm diameter PFL, designed to image at 8 keV, as compared to the design profile (dashed).  The top plot shows the results at the center of the lens while the bottom is that for near the outermost region of the lens.}
  \label{prof_fig}
 \end{figure}
 \noindent
The gray-scale lithographic technique was used to fabricate silicon PFLs of substantial diameters in a format scalable for astronomical implementation, with micron minimum feature size and focal length appropriate for ground tests. The gray-scale mask design and lithography were performed at the MSAL at the University of Maryland\cite{Morgan1}: an optical mask was generated with the desired structure built from small, varying opacity pixels. By modulating the intensity of light through a gray-scale optical mask, a positive photoresist that was spun onto a silicon substrate was partially exposed to different depths. After development, a 3-dimensional profile made of Ôgray levelsÕ remains in the photoresist corresponding to the intensity pattern generated by the optical mask. The structure was then transferred into the silicon via an anisotropic, Deep-Reactive-Ion-Etch (DRIE) to fabricate the desired device.

Two sets of silicon PFLs have been fabricated: one set designed for 8 keV and the other set for 17.4 keV.  The fabricated devices have been characterized via metrology, such as optical profilometry, as well as by X-ray  measurements of their imaging performance. The PFLs designed to image at 8 keV (Cu K$\alpha$) have a focal length of 113 meters. A SEM of a this 3 mm diameter PFL is shown in Figure \ref{pfl_fig} and results obtained from profilometry are presented in Figure \ref{prof_fig}. The PFLs designed to image at 17.4 keV (Mo K$\alpha$) have a focal length of 110 meters.  The 8 keV (17.4 keV) PFLs were fabricated with a 3 mm diameter over a 30 $\mu$m (60 $\mu$m) thick substrate and with a Fresnel ridge height of $2 \times t_{2\pi} \approx$ 40 $\mu$m (90 $\mu$m). These $4\pi$ PFLs have each Fresnel ridge profile spanning 2 full Fresnel zones and designed to have 16 individual steps forming each Fresnel ridge profile.  The minimum Fresnel ridge spacing, at the outermost radii, is 24 $\mu$m for the 8 keV PFL and 10 $\mu$m for the 17.4 keV PFL.  4.7 mm diameter 8 keV PFLs have also been fabricated as well as 2$\pi$ versions of the 3 mm diameter 17.4 keV PFLs.  Table \ref{table_PFL} compares some of the design parameters between the 8 and 17.4 keV PFLs.
 \begin{table}[!h]
  \caption{PFL Parameters: 4$\pi$ Design}
  \label{table_PFL}
  \centering
 \begin{tabular}{cccc}
\hline
PFL Design Energy & Diameter & $p_{min}$ & N$_{Ridges}$   \\
\hline
8 keV & 3 $mm$ & 24 $\mu m$ & 32 \\
17.3 keV & 3 $mm$ & 10 $\mu m$ & 71  \\
\hline
\end{tabular}
  \end{table}
\section{Experimental Results}
The characterization of the imaging properties of the fabricated PFLs was performed at the NASA GSFC 600-meter X-ray Interferometry Testbed.  The PFL under test was placed approximately 146 meters from a $\mu$focus x-ray source of appropriate energy, and a LN$_2$-cooled CCD x-ray camera was located approximately 453 meters downstream from the PFL. The effective focal length of this arrangement is 110.4 m. The 35 $\mu$m thick CCD contained $1024 \times 1024$, $13 \times 13$ $\mu$m$^2$ pixels and demonstrated an energy resolution of 150 eV (FWHM) at 8 keV (Cu K$\alpha$).  The 600 m beam line and chamber containing the PFLs were held at a $\sim$1 mTorr pressure during testing.  A narrow tungsten slit waw employed to determine the point-spread-function of each tested optic with the slit located a few cm from the x-ray source, thus constraining the x-ray spot size in one dimension. We have characterized three different PFLs all with 3 mm diameters: a PFL designed for 8 keV, a PFL designed for 17.4 keV, and a contact-pair achromat.  
\begin{figure}[!t]
  \centering
  \includegraphics[width=2.5in]{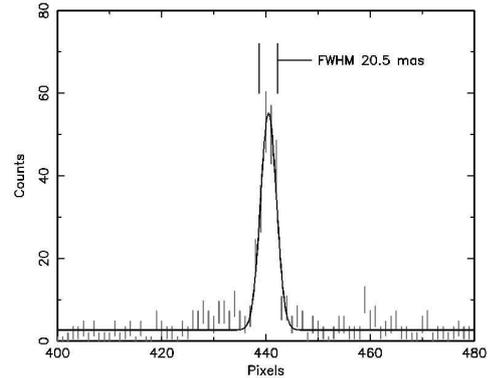}
  \caption{Projection of the image of a 5 $\mu$m slit formed in Cu K$\alpha$  x-rays (8.04 keV)  with a  3 mm diameter  PFL. The focus has been trimmed with a refractive corrector lens. For comparison, the diffraction-limited response would correspond to 16.7 mas (m$^{\prime\prime}$) which includes slit magnification and detector resolution effects.} 
  \label{diflim_fig}
 \end{figure}
The 8 keV PFLs were characterized with a Cu-target X-ray tube and measured to have efficiencies of $\sim$35\% at their design energy of 8 keV. This is 70\% of the theoretical maximum of $\sim$50\% for this PFL which includes the effect of absorption from the substrate and $4\pi$ Fresnel profile. The loss in flux measured in the peaks is consistent with that observed in a circular interference pattern, located around the central peak, that two independent simulations determined to be caused by fabrication artifacts and the 4$\pi$ PFL implementation.
\begin{figure*}[!t]
   \centerline{\subfloat[8.4 keV]{\includegraphics[width=1.9in]{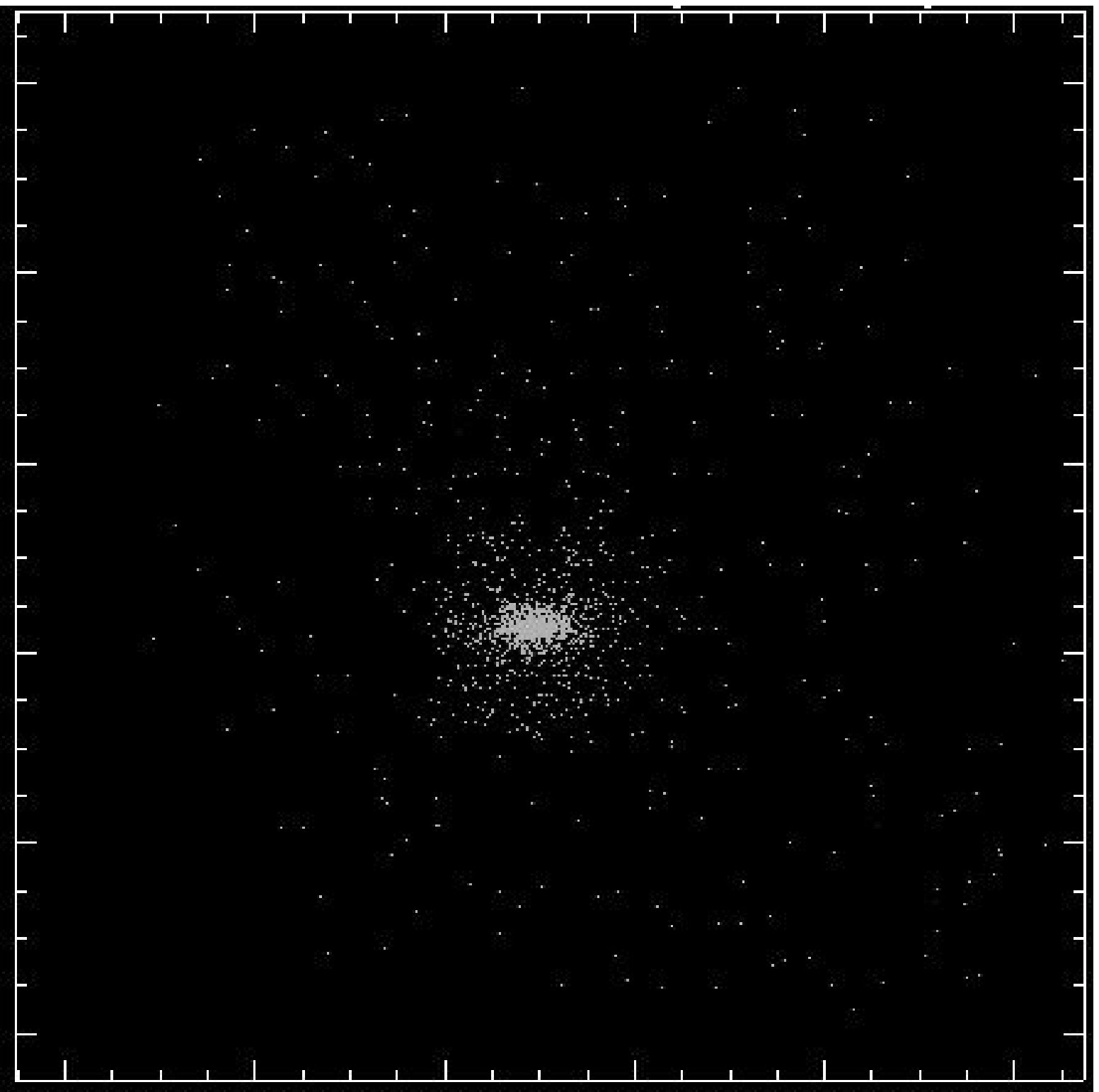} \label{sub_fig1}}
              \hfil
              \subfloat[9.7 keV]{\includegraphics[width=1.9in]{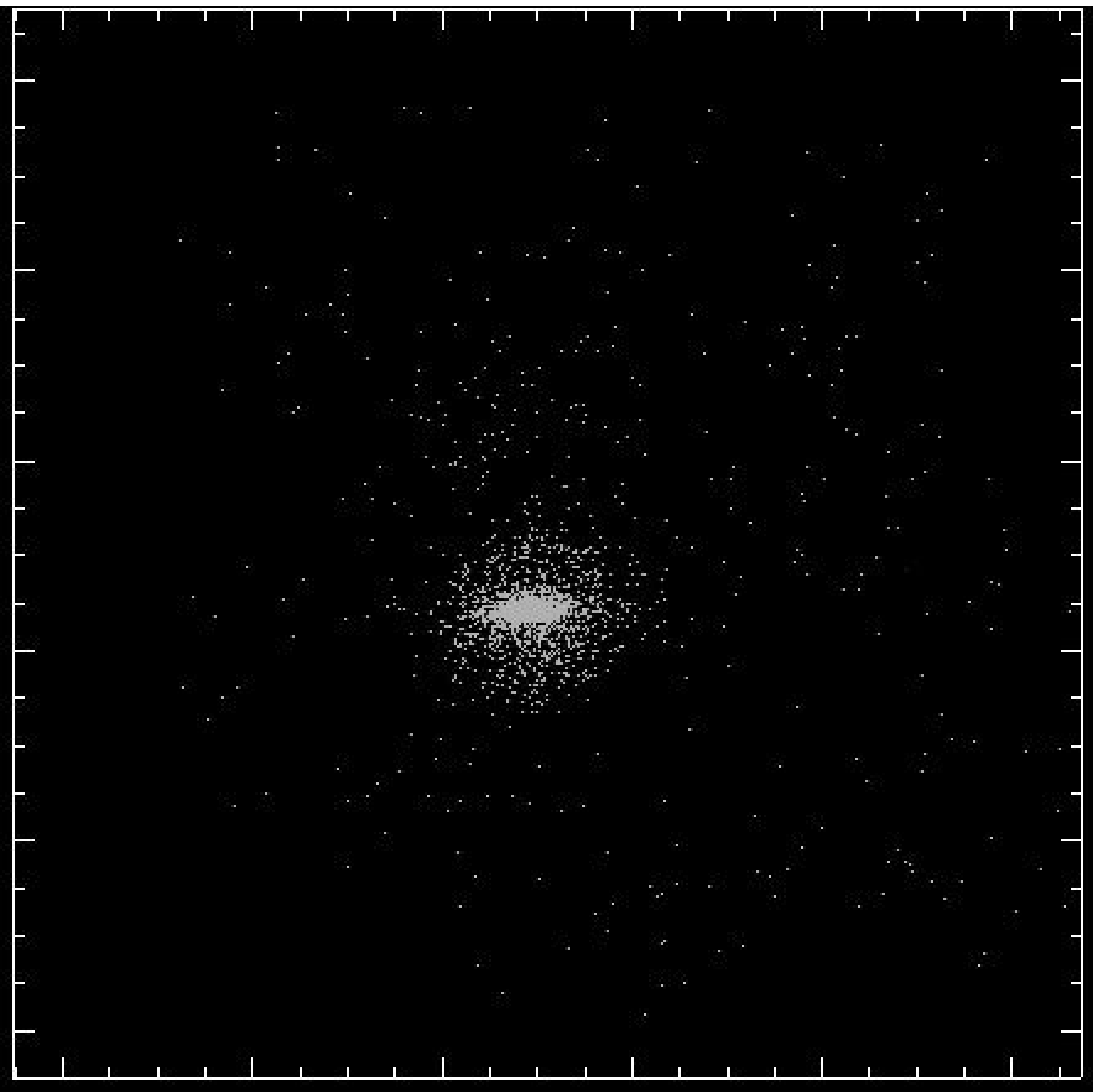} \label{sub_fig2}}
                            \hfil
              \subfloat[11.3 keV]{\includegraphics[width=1.9in]{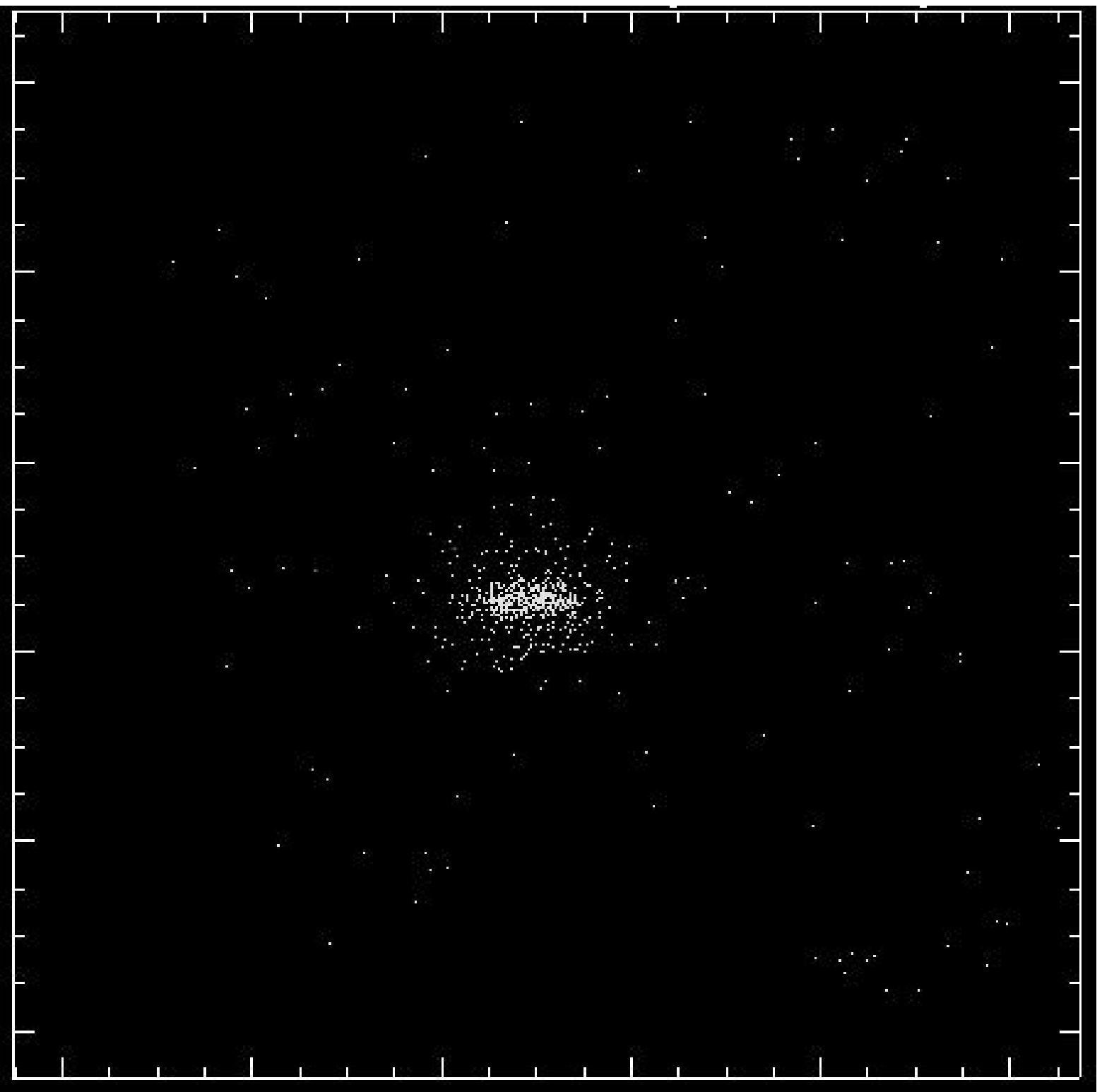} \label{sub_fig3}}
             }
   \caption{CCD images of a 5 $\mu$m wide slit as imaged by a refractive-PFL achromat at 8.4, 9.7, and 11.3 keV using the L-lines from a tungsten $\mu$focus x-ray tube.  The images show consistent widths independent of energy.  The spot size would encompass an entire frame at 11.3 keV for a simple PFL designed to image at 8.4 keV.}
   \label{achromat_fig}
 \end{figure*}
The focal length of the 8 keV PFLs is slightly longer than the 110.4 m needed for the 600 m test beam configuration.   Furthermore, the source-PFL distance (146 m) and PFL-detector distance (453 m) used in the test beam effectively magnifies any focal length differences in terms of image broadening as measured by the detector.  The focal length of the PFL was corrected to that needed for the test beam arrangement by incorporating a refractive corrector lens with a PFL to form a compound optic. The power of the refractive component, fabricated using diamond-turned Zeonex, was chosen to bring the 8 keV, Cu K$\alpha$ photons into focus, i.e. change the focal length of the refractive-diffractive combination to 110.4 m from the PFLÕs focal length of 113 m. The results are shown in Figure \ref{diflim_fig} and demonstrate that the inherent imaging of the PFL is nearly diffraction-limited, corresponding to an angular resolution of 20.5 m$^{\prime\prime}$. Initial measurements using a 17.4 keV PFL, which has a focal length more closely matched to the test beam configuration, have demonstrated imaging within a factor of 4 of the diffraction limit.

In order to experimentally demonstrate the imaging performance of a PFL-based achromat, a contact-pair achromat was constructed by combined a Zeonex parabolic refractive lens, fabricated with diamond turning, and a PFL component fabricated using the 17.4 keV design. This exploited a property of PFLs: a PFL designed for an energy $E$ will image at an energy $E^\prime$ with a focal length $f^\prime = (E^\prime/E) f$.  The PFL component was fabricated with a Fresnel ridge height ($\sim$20 $\mu$m) to improve efficiency at $\sim$8 keV and has a focal length approximately half of that at 17.4 keV, which is needed for an achromat operating $\sim$8 keV.  The thickness profile of the refractive component is given by $t(r)=r^2/(2f\delta)$ where $r$ is the radial dimension of the lens, $f$ is the focal length, and $\delta$ is the refractive index decrement.  The height of $\sim$5 mm is to be compared to the attenuation length of 2.9 mm at 8.4 keV. While the tested achromat was not optimized for efficiency, it was used to demonstrate that achromatic imaging over a range of x-ray energies can be achieved. We have performed initial characterization on this refractive-diffractive achromat using the L-lines of tungsten (8.4, 9.7, 11.3 keV) by imaging a 5$\mu$m slit. The CCD used in the 600 m beamline has sufficient energy resolution to separate the W L-lines.  Figure \ref{achromat_fig} shows images of a slit source obtained using the achromat at the three energies and demonstrates consistent, achromatic imaging over a wide energy range. The image at 11.3 keV from a simple PFL designed for 8.4 keV (with $f=110.4$ m) would fill an entire image frame, as implied by Equation \ref{focal_eqn}.
\section{Conclusions}  
We have fabricated prototype, silicon PFLs, using the MEMS gray-scale technology, and have characterized their imaging properties in a 600 meter x-ray beam line. The results demonstrate near diffraction-limited imaging with high collection efficiency at 8 keV.  We have also demonstrated that the chromaticity of PFLs can be alleviated with the introduction of a refractive component to form an achromat.  These results demonstrate the superior imaging potential in the X-ray/$\gamma$-ray energy band for PFL-based optics employed in astronomical instrumentation. This research has been supported under NASA NNH04ZSS001N-APRA, NNH06ZDA001N-APRA, and NNH07ZDA001N-APRA.

\end{document}